\documentclass[10pt,journal,compsoc]{IEEEtran}

\usepackage{comment}

\ifCLASSOPTIONcompsoc
  \usepackage[nocompress]{cite}
\else
  \usepackage{cite}
\fi
\usepackage{hyperref}
\usepackage{amsmath}
\ifCLASSINFOpdf
\else
\fi
\def\BibTeX{{\rm B\kern-.05em{\sc i\kern-.025em b}\kern-.08emT\kern-.1667em\lower.7ex\hbox{E}\kern-.125emX}}
\usepackage{float}
\usepackage{graphicx}
\usepackage{subcaption}
\usepackage{cleveref}
   \usepackage{wrapfig}
\usepackage{caption}

\usepackage[left=0.56in, right=0.56in, top=0.75in, textheight=695pt]{geometry}

\begin{document}
%
\title{Resilience-by-design in Adaptive Multi-Agent Traffic Control Systems}
%
%
%
%

\author{Ranwa~Al~Mallah,~\IEEEmembership{Member,~IEEE,}
        Talal~Halabi,~\IEEEmembership{Member,~IEEE,}
        and~Bilal~Farooq,~\IEEEmembership{Member,~IEEE}
\IEEEcompsocitemizethanks{\IEEEcompsocthanksitem Ranwa Al Mallah is with the Computer Security Laboratory in the Electrical and Computer Engineering Department, Royal Military College of Canada, Kingston K7K 7B4, ON.  \protect \hfil\break
E-mail: ranwa.al-mallah@rmc-cmr.ca. \protect \hfil\break
\IEEEcompsocthanksitem Talal Halabi is with the Department of Applied Computer Science, the University of Winnipeg, Winnipeg R3B 2E9, MB.\protect \hfil\break
E-mail: t.halabi@uwinnipeg.ca. \protect \hfil\break
\IEEEcompsocthanksitem  Bilal Farooq is with the Laboratory of Innovations in Transportation at Ryerson University, Toronto M5B 2K3, ON.\protect \hfil\break
E-mail: bilal.farooq@ryerson.ca.
}}

%
%

{}
%



\IEEEtitleabstractindextext{%
\begin{abstract}
Connected and Autonomous Vehicles (CAVs) with their evolving data gathering capabilities will play a significant role in road safety and efficiency applications supported by Intelligent Transport Systems (ITS), such as Traffic Signal Control (TSC) for urban traffic congestion management. However, their involvement will expand the space of security vulnerabilities and create larger threat vectors. In this paper, we perform the first detailed security analysis and implementation of a new cyber-physical attack category carried out by the network of CAVs against Adaptive Multi-Agent Traffic Signal Control (AMATSC), namely, coordinated Sybil attacks, where vehicles with forged or fake identities try to alter the data collected by the AMATSC algorithms to sabotage their decisions. Consequently, a novel, game-theoretic mitigation approach at the application layer is proposed to minimize the impact of such sophisticated data corruption attacks. The devised minimax game model enables the AMATSC algorithm to generate optimal decisions under a suspected attack, improving its resilience. Extensive experimentation is performed on a traffic dataset provided by the City of Montr\'eal under real-world intersection settings to evaluate the attack impact. Our results improved time loss on attacked intersections by approximately 48.9\%. Substantial benefits can be gained from the mitigation, yielding more robust adaptive control of traffic across networked intersections.
\end{abstract}

}

\maketitle

\IEEEdisplaynontitleabstractindextext

%
\IEEEpeerreviewmaketitle

\section{Introduction}
Intelligent Transportation Systems (ITS) play a vital role in the development of smart cities by enabling a plethora of road safety and efficiency applications such as optimized traffic management and collision avoidance. Adaptive Multi-Agent Traffic Signal Control (AMATSC) is one of the aspects that make these systems intelligent by increasing the responsiveness of traffic signals to traffic demands. The cost of traffic congestion in wasted fuel and lost productivity ranges between 1.5 and 5 billion CAD per year in major Canadian cities \cite{stats}. By collecting and evaluating traffic data in real time, AMATSC technologies try to optimize signal timings to reduce urban congestion and ensure reliable vehicle travel time. 

\indent AMATSC are deployed extensively in North America, Australia, and Europe. For instance, the Sydney Coordinated Adaptive Traffic System (SCATS), Split Cycle Offset Optimization Technique (SCOOT), RapidFlow, Rhythm engineering, InSync, Urban Traffic OPtimisation by Integrated Automation (UTOPIA), Adaptive Control Software (ACS Lite), and Real-Time Hierarchical Optimized Distributed and Effective System (RHODES) represent widely known deployments \cite{studer2015analysis}. They are in use in Pittsburgh, Pennsylvania, on Boudreau Road in the City of St. Albert, Alberta, Canada, in the suburb of Toronto, the Holton region and in downtown Toronto. By the end of 2021, the size of the deployment is expected to grow to hundreds more in operation in downtown Toronto, Montreal, and many other cities around the world \cite{smith2013smart}. 

\indent The network of Connected and Autonomous Vehicles (CAVs) will be increasingly involved in AMATSC operations, where valuable traffic parameters that are often difficult to obtain from the static transportation infrastructure are transmitted through the vehicles' On Board Units (OBU) to the infrastructure's Road Side Units (RSU) and integrated into the AMATSC algorithms to optimize their signal timing decisions. However, the reliance on CAV increases the vulnerability of ITS to cyberattacks \cite{checkoway2011comprehensive} due to the high connectivity involved. AMATSC may be implemented and managed by the Traffic Management Center (TMC). Microscopic and macroscopic traffic variables collected at the RSUs and transmitted to the TMC are critical to AMATSC algorithms for optimal decision making. However, they may be misleadingly altered in the presence of malicious vehicles trying to launch data corruption attacks, such as Sybil. Basically, Sybil vehicles are non-existing vehicles controlled by a malicious entity and claiming fake or forged identities to participate in ITS operations, sabotaging their reliability and performance. 

\indent The communication protocols used by traffic cabinets lack effective data encryption and authentication mechanisms \cite{zetter2014hackers}. Consequently, the Sybil attack exploits the lack of security countermeasures on traffic controllers and sensors to expose the vulnerabilities of AMATSC algorithms. Sybil attacks in Peer-to-Peer (P2P) networks have been well studied in the literature, and many solutions were designed to identify and isolate Sybil nodes within these networks \cite{levine2006survey}. However, Sybil attack prevention and detection in wireless networks is not straightforward when the attack becomes complex. Hence, it is imperative to equally focus on designing effective mitigation solutions to preserve the resilience of traffic applications when the primary lines of defense are compromised.\\
\indent In this paper, we investigate theoretically and experimentally the potential and impact of a yet unexplored threat model on AMATSC systems, namely the coordinated Sybil attack. In such an attack, the Sybil vehicles are deployed to alter the data collected by the AMATSC algorithms on networked intersections by optimally targeting the controllers involved in signal timing decisions. The motivation behind our security analysis is to stress the need for these algorithms to be attack-aware. This could be achieved through the integration of mitigation strategies into the AMATSC decision making as a resilience layer against sophisticated attacks. 

Therefore, we devise a non-cooperative minimax game model to formalize the interactions between the attacker and the AMATSC system, then solves the game to generate the optimal attack and defense strategies. On one hand, the coordinated Sybil attack consists in playing a mixed strategy drawn from the attack action space and is reflected by the adaptive deployment of Sybil vehicles on networked road junctions to maximize their traffic flows. On the other hand, the AMATSC algorithm adopts a mixed strategy of attack mitigation actions that consists of applying a weighted integration of the data collected by the network of traffic controllers in an attempt to optimize signal timing decisions under attack.\\
\indent The contributions of this paper are summarized as follows:
\begin{itemize}
\item We perform the first detailed security analysis of a highly realistic cyber-physical threat model carried out by a network of CAVs on AMATSC systems. We experimentally demonstrate its effectiveness and measure its impact on traffic control decisions using a real traffic dataset provided by the City of Montr\'eal. \vspace{2mm}
\item We propose a novel, application-layer Sybil attack mitigation approach based on a minimax game model. Unlike other mitigation solutions that focus on reducing the impact of the attack following its detection, our approach integrates a resilient response to attacks into the generation of decisions. This 'by-design' solution can be considered as a fail-safe of the attack detection phase and can be deployed as the default operational strategy under susceptible adversarial settings where critical traffic data might be maliciously corrupted.\vspace{2mm}
\item We implement the mitigation approach onto an existing Multi-Agent Reinforcement Learning (MARL)-based AMATSC algorithm and evaluate its performance under various scenarios. The results show that attack mitigation substantially reduces vehicle delays and yields optimal control of traffic across networked intersections under attack.
\end{itemize}\vspace{1mm}

\indent The remainder of the paper is organized as follows. Section \ref{sec:1} presents important background concepts and discusses the related literature. Section \ref{sec:2} describes the new threat model and its impact. Section \ref{sec:3} lays out the proposed mitigation approach. The implementation details are presented in Section \ref{sec:4}. Section \ref{sec:5} discusses the experimental setup and analyzes the results. Finally, Section \ref{sec:6} concludes the paper.

\section{Background and Literature Review}
\label{sec:1}

\indent The dependability and safety of ITS rely on its security against cyber and cyber-physical attacks, which target the perception layer of the architecture and propagate their impact to the application layer.  Traffic Signal Control (TSC) has evolved from standalone hardware devices running static schedules into complex, wireless connected systems, which exposed them to a variety of cyberthreats \cite{studer2015analysis}. 
Signal controllers are usually placed in metal cabinets by the roadside and used to generate signal timing plans by following different approaches: fixed-time, actuated and semi-actuated, and adaptive. In particular, adaptive TSC has proven to increase transportation productivity and reduce gas emissions.

\begin{figure}[!t]
	\includegraphics[width=0.45\textwidth]{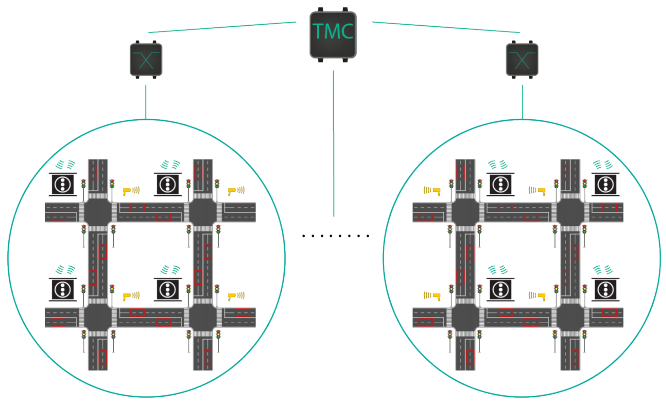}
	\centering
	\caption{Overview of AMATSC system architecture.}
       \label{figure:ATSC}\vspace{-2mm}
\end{figure}

\subsection{Adaptive Multi-Agent TSC}
AMATSC emerged as a new ITS paradigm, persuaded by the idea that traffic light timing plans should be fully adaptive in coping with actual traffic demand (intersections' traffic load). Rather than monitoring the states of isolated intersections, AMATSC observes the state of networked intersections to produce signal timing decisions based on the global state of the road network \cite{dusparic2016towards}. To achieve this, the local controllers at every intersection communicate with each other directly or via the TMC, as depicted in the architecture of Figure \ref{figure:ATSC}. \\
\indent The potential for performance improvements in AMATSC driven by Reinforcement Learning (RL) when compared to conventional approaches is very promising \cite{mannion2016experimental}. 
When developing MARL-based AMATSC models, the literature draws on different parameters for state and reward definitions within the Markov game played among the agents \cite{bazzan2009opportunities}. For instance, queue length is often used to define the environment state. Some studies propose a vehicle-based state definition using the expected total waiting time of a vehicle before reaching its destination. More generally, delay-based approaches exploit a combination of parameters such as the queue length and the traffic flow rate, or the queue length and the time elapsed since the previous signal. Similarly, a variety of objectives have been considered when estimating the reward function such as the average trip time, average junction waiting time, and junction throughput/flow rate.

\subsection{Security of AMATSC}
\begin{figure}[!t]
	\includegraphics[width=0.4\textwidth]{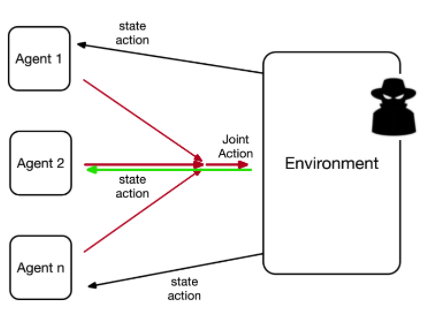}
	\centering
	\caption{Attackers can distort the environment states and agents' rewards, leading to suboptimal MARL policies.}       \label{marlattack}\vspace{-2mm}
\end{figure}

AMATSC algorithms manage the signal timings of multiple networked intersections. CAV-based traffic control is the future of AMATSC and will greatly improve its efficiency by allowing the vehicles to transmit relevant information to urban traffic scheduling algorithms. However, this increased connectivity aggravates the vulnerability of AMATSC algorithms to falsified data and erroneous measurements, especially if they are not designed with the intrinsic ability to detect or ignore these measurements during decision making. For instance, Fig. \ref{marlattack} shows how a MARL-based AMATSC can be targeted by data corruption attacks that may sabotage the environment conditions on which the traffic control agents rely to derive their optimal signal timing policy. \\
\indent Ghena et al. \cite{ghena2014green} investigated the security of an AMATSC system currently deployed in the United States and discovered a number of serious security flaws, which mainly exist due to systemic design failures and may be exploited to create attacks that gain control of the system and cause service disruption. Chen et al. \cite{chen2018exposing} conducted a detailed security analysis of CAV-based transportation and studied the impact of data spoofing attacks performed by the vehicles. They found that the current traffic control algorithm design choices are highly vulnerable to such attacks from even a single vehicle. Yen et al. \cite{yen2018falsified} also studied the performance of different AMATSC scheduling algorithms when they are under attack. Feng et al. \cite{feng2018vulnerability} focused on attacking actuated and adaptive control systems by sending falsified data to increase traffic delay. Although these attacks are still relatively simple, they attempt to emphasize the vulnerability of scheduling algorithms.\\
\indent Laszka et al. \cite{laszka2019detection} proposed an approach for detecting and mitigating the attacks on traffic signals by optimizing the detector's configuration, which implies that the mitigation process is based upon the success of the detection phase. Our work underlines the importance of deploying the mitigation procedure at the application layer during the design phase without relying on the precedent line of defense, where the AMATSC algorithm integrates an intrinsic form of smart resilience when processing the data collected from the physical world independently of intrusion detection functions.

\subsection{Resilience-by-design}
\indent To the best of our knowledge, the robustness of the parameters used for state and reward definitions in the face of cyberattacks against the ITS remains questionable. The attack carried out in this paper proves that the intrinsic parameters used by the traffic control logic are extremely vulnerable to corruption and require serious investigation. Thus, we stress the need to develop resilient-by-design AMATSC algorithms as part of a defense-in-depth security approach, where an additional mitigation layer is deployed on top of the application layer. 

The idea of security-by-design for cyber-physical systems has been introduced very recently \cite{geismann2018towards}. In principle, it allows to incorporate security measures at the system design phase following careful specification of security requirements. This paper largely contributes to this area of research, which is still in its infancy, by emphasizing the need to equally introduce security and resilience-by-design at the application layer of the cyber-physical architecture. This is a crucial line of defense, especially against intelligent attacks and advanced persistent threats that will most likely bypass detection.

Game theory provides a high-level mathematical language, generally perceived as effective in modeling the conflicted interactions between attackers and defenders in modern computer systems and optimizing their decisions. Several games have been proposed to detect the attacks on ITS and address traffic prediction and decision-making issues \cite{sedjelmaci2019cyber}. In such games, usually non-cooperative, players aim at maximizing their own payoffs and minimizing the payoffs of their opponents \cite{alpcan2011security}. The attacker's payoff can be modeled either as the gain from launching the attack on the ITS or as the impact of the attack on system performance. In this paper, we leverage the power of game theory to anticipate the malicious activities that target ATSC and prepare the controllers to respond proactively and in an optimized fashion. 
%
	
\subsection{Sybil Attacks in the CAV Environment}
Sybil attacks were first introduced by Douceur \cite{douceur2002sybil} in P2P networks: If a single faulty or malicious entity can present multiple identities, it can control a substantial fraction of the system, thus representing a security threat. Many Sybil attack detection techniques were proposed \cite{vasudeva2018survey}, but most of them present serious shortcomings in practicality and detection accuracy, especially if faced with sophisticated attacks. In fact, without a centralized trusted authority, it is impossible to completely detect and eliminate Sybil nodes \cite{douceur2002sybil}. Nonetheless, a centralized trusted authority is not necessarily feasible in vehicular networks, where vehicle mobility and the dynamic network topology substantially hinder the design assumptions behind most attack detection schemes \cite{vasudeva2018survey}. 

\indent In vehicular networks, there is an urgent need to explore attack mitigation strategies as it was done in wired P2P networks in order to minimize the malicious effects of the Sybil attack. For example, one mitigation strategy is to disrupt the activity of the Sybil node. However, this is not directly applicable to vehicular networks, because in the case of a Sybil attack with multiple Sybil identities, the strategy must detect multiple malicious nodes consequently, thus degrading the communication and performance of the vehicular network. Hence, the attenuation of the attack impact on the performance of the network remains a critical defense aspect.

\indent This paper addresses the gap of enacting the right security posture in ITS environments susceptible to Sybil attacks, particularly their evolving versions, which strategically target multiple traffic controllers to exacerbate the interference with the decisions of traffic flow management applications overseeing networked intersections. These attacks will gain control over a significant part of the network to influence unfavorably the TMC in various network operations. An attacker with multiple fake identities can maliciously participate in a data aggregation process, take advantage of resource allocation mechanisms, disrupt vote-based mechanisms, and interfere with traffic flow management and navigation applications. A defense system assuming a centralized trusted authority is less likely to be feasible in the ITS. 

%


\section{Threat Model}
\label{sec:2}
ITS reliability is highly dependent on the quality of data collected across the system. This section describes a yet unexplored threat model that targets data integrity in AMATSC.

\subsection{Coordinated Data Corruption: The Case for Sybil}
In normal settings, the vehicles send a message to indicate their position and arrival time when approaching a traffic controller. Hence, the latter is fully aware of the number of vehicles in each lane. To generate signal-timing decisions, the queue-based control algorithm will take into consideration the perceived number of vehicles as the queue length, while the delay-based algorithm will extract the vehicles' delay at the intersection from their broadcasted arrival times. In MARL-based AMATSC, these information will be used in the stochastic game to estimate the transition probabilities between the different states. Hence, corrupting these information will push the controllers off the optimal policy distribution. \\
\indent A Sybil attack on traffic controllers consists in creating or forging fake vehicle identities and using them to broadcast traffic-related information that may compromise the decisions of traffic flow management applications. This attack is considered indirect since it does not require to physically access the traffic controller, but to tamper with data transmitted by the network of CAVs, which makes it highly feasible in reality. This paper explores an evolved version of Sybil attack, which we call coordinated Sybil attack, that will mainly target highly interconnected systems. In this scenario, the attacker will make use of fake vehicle identities to simultaneously target multiple traffic controllers by distributively corrupting their data inputs with the intent of producing a larger attack impact. This can substantially degrade the performance of ITS applications that rely on multiple networked data sources to generate decisions and take actions. For MARL-based multi-agent systems, this threat can be seen as a direct distortion of the environment on which the agents rely to compute their rewards and derive their policies.

\subsection{Threat Impact}
\indent By optimally dispersing the attack load, the attacker is not only motivated by the greater scale and consequence of the attack, but also seeking to hide it from anomaly-based intrusion detection systems that may raise security alarms if the attack is concentrated onto a single controller. Also, the number of fake identities generated by the attack onto a single traffic lane should adhere to the traffic flow capacity of the lane to avoid raising an intrusion flag. The attack assumes the ability of the malicious vehicle to compromise its OBU and transmit malicious messages. This can be performed physically, wirelessly, or via malware \cite{mazloom2016security}. \\
\indent The impact of coordinated data corruption attacks increases substantially when simultaneously targeting  multiple networked intersections controlled by an AMATSC algorithm, as proven experimentally later in this paper. For instance, Sybil vehicles within the network are simply regarded as fake messages, but the RSU deploying AMATSC functions actually receives these messages. Thus, the traffic light state may change to discharge traffic that is not really present. Hence, the lanes carrying the actual traffic can become extremely congested because the presence of Sybil vehicles altered the data received by the MARL agents and used by the algorithm in the optimization of timing plans.


%
\subsection{Attack Assumptions}
In the coordinated Sybil attack, the attacker's goal is to maximize the impact on decision making while minimizing the probability of detection. Hence, the attacker will deploy the Sybil nodes under strategically placed RSUs, the ones that the attacker thinks they will influence the decision of the TMC the most according to the perceived delay. The attack can be driven by political or financial incentives, or may be carried out to cause damage to city functions and individuals (e.g., terrorism). We assume that the attacker has sufficient resources and incentives to monitor the traffic on the target networked intersections for long periods of time. The attacker also has full knowledge of the intersection map. Hence, before attacking an intersection, the attacker is assumed to have performed sufficient reconnaissance and thus have studied beforehand the appropriate timing to launch the attack in order to create the most impact, which could be specific to the target junction (e.g., targeting rush hour). Also, the main attack vehicle does not have to be present in the traffic flow, but could be launching the attack from a distance. \\
\indent We assume that the attacker is only interested in manipulating the number of vehicles in a single direction, thus deploying fake identities on opposite directions is not useful. The attacker can define the critical signal phases to target for increased impact. Although trying out all attack scenarios guarantees the optimal attack strategy, it is unrealistic for an attacker to enumerate them all and find out the best strategy in real time based on the observed attack impact, which needs to be effectively estimated. Thus, a simple greedy attack policy is usually proposed to find an effective attack strategy that yields a sufficient impact according to the attacker. Nevertheless, our proposed game model will enable the attacker to apply an optimal attack policy by solving an optimization problem that produces the best-case attack scenario. 

\section{Attack Mitigation: A Minimax Game}
\label{sec:3}
\indent  This paper proposes a coordinated Sybil attack mitigation scheme as a resilience-by-design approach to be integrated into the decision-making phase of AMATSC systems, providing real-time, dynamic protection in adversarial scenarios.

\subsection{Game Overview}
\indent Game Theory has been widely applied to solve a variety of security problems in computer and communication networks. In this paper, we design a minimax game theoretical model to describe the interactions between the AMATSC algorithm and the attacker. By solving the game, the algorithm determines the optimal defense strategy to deploy in order to minimize the impact of the attack in the worst-case scenario. 

The data corruption attacker can implement a mixed strategy of attack actions by targeting the traffic intersections with different magnitudes based on the estimated potential impact. This also allows the attacker to mask the effects of the attack while prolonging its duration. The payoff of the attack is quantified in terms of the impact on traffic flow at the target intersections. On the other hand, the AMATSC algorithm is not aware of the intersections being targeted by the attack, hence it must estimate the impact based on the status of the traffic network and consider the information received by each controller accordingly, even though this might entail a cost associated to disregarding some legitimate vehicle messages. The resilience layer proposed in this paper is agnostic to the attack strategy and aims to prepare the system to function in the worst adversarial setting without prior knowledge of attacked traffic controllers. 

The proposed game is a zero-sum game in which one player's payoff is the negation of the other player's payoff. The game is simultaneous, since the defense strategy is implemented regardless of the attack status and is considered as an additional security layer, which is independent from the attack detection functions in place. This recursively played game can be considered as a single-state stochastic game, which will be solved by stationary strategies that do not depend on history and time slots.

\subsection{Attack and Defense Strategies}
\indent In the cybersecurity battle between the attacker and traffic control algorithm, a set of attack actions is represented by $\mathcal{A}=\{a_1,\dots,a_j,\dots,a_A\}$, where $A$ is the size of the attacker's action profile. The pure strategy $a_j$ consists in targeting lane $j$ each time. The mixed strategy in this case consists in defining a probability distribution vector denoted by $\alpha=(\alpha_1,\dots,\alpha_A)$, such that $\alpha_j \geq 0\ \forall a_j \in \mathcal{A}$ and $\sum \limits_{a_j\in \mathcal{A}}\alpha_j=1$, to attack the network of intersections with different rates of Sybil vehicles. It is intuitive to assume that the size of the action profile is equal to the total number of traffic lanes at all target intersections. \\
\indent The TMC controlling the AMATSC algorithm implements the mitigation approach through a set of actions represented in the set $\mathcal{D}=\{d_1,\dots,d_i,\dots,d_D\}$, where $D$ is the size of the defender's action profile. As in the case of attack actions, the defense actions will also be applied at the lane granularity level, assuming that each lane is governed by a distinct traffic controller. Hence, it is intuitive to assume that the size of the attack and defense action profiles is the same. 

The AMATSC algorithm usually takes the calculated average speed of vehicles present at the traffic light as input and generates the signal timings accordingly. A pure strategy $d_i$ consists in considering all the vehicles in lane $i$ when computing the input average speed. Adopting a mixed strategy of defense that mitigates the attack impact consists in modifying the input vector to the AMATSC algorithm over all lanes according to a probability distribution vector produced by the solution of the game over the actions in $\mathcal{D}$. The vector is denoted by $\beta=(\beta_1,\dots,\beta_D)$, such that $\beta_i \geq 0\ \forall d_i \in \mathcal{D}$ and $\sum \limits_{d_i\in \mathcal{D}}\beta_i=1$. Our mitigation solution aims to find the optimal vector $\beta$ and integrate it into the design of the traffic control logic to reduce the coordinated Sybil attack impact. This is one of the innovative ideas advocated in this paper.

\subsection{Payoff Matrix}
\indent The payoff of the attacker can be determined based on the type of the data corruption attack and the caused impact. In the case of coordinated Sybil attacks, the payoff of the attacker is equal to the difference between the maximal flow on the traffic lane, denoted by $\theta_i$, and the actual traffic flow on that lane, denoted by $f_i$. These parameters are particularly chosen as they allow to effectively estimate the attack impact. The utility matrix created in our game model is a square matrix of size $D$, which is equal to the total number of segments on the networked intersections. Each row corresponds to a defense action that may be adopted by the AMATSC algorithm, which in this case represents the lane selected by the traffic controller and whose data is fed as input to the scheduling algorithm to generate the signal timing plan. On the other hand, each column corresponds to an attack action, which mainly represents the target traffic lane. \\
\indent The attacker must maximize the attack impact on the decisions of the AMATSC algorithm by attempting to corrupt the collection of input traffic data from the lanes as unfavorably as possible. Thus, the utility matrix $U$ of the attacker, which is used to compute the attacker's payoff over the set of possible strategies, can be defined in terms of the attack impact by:
\begin{equation}\label{payoff}
U_{ij} = \left\{\begin{matrix}
 \theta_i-f_i, & i=j\\
0, & i\neq j
\end{matrix}\right.
\end{equation}
The impact is equal to 0 when the AMATSC algorithm does not consider the target lane in its decision making. On the other hand, when $i=j$, the algorithm takes into account only the traffic data of lane $i$ to generate signal timing decisions (i.e., pure strategy) while the attacker targets the same lane using the Sybil vehicles. Hence, the attack impact is maximal, because the input data is completely corrupted and was perceived by the algorithm as fully accurate. \\
\indent However, the AMATSC algorithm does not intend to deploy pure strategies, since its main function consists in observing the states on all intersections involved. The goal is then to apply the mitigation procedure on the input data vector itself before feeding it to the algorithm. Hence, the mixed strategy is reflected by the assignment of distinct degrees of confidence in the perceived traffic parameters on all lanes according to the probability distribution vector $\beta$ generated by the game. For example, a value $\beta_1=0.6$ will dictate that the algorithm only considers 60\% of the vehicles present on lane $1$ when computing the associated input data instead of 100\% of the vehicles, in an attempt to anticipate the load of Sybil nodes on that lane and reduce its effect. The definition of this utility matrix leads to an AMATSC vs. attacker zero-sum game, where the payoff of the algorithm is equal to the negation of the payoff of the attacker. In other words, the algorithm would lose what the attacker would gain, and vice versa.

\subsection{Game Model}
Based on the above definitions of attack and defense action spaces as well as payoff matrices, we propose a two-players zero-sum minimax game, in which the objective of the attacker is to distribute the Sybil attack load over the set of traffic segments to maximize the attack impact, and the objective of the traffic control algorithm is to determine the optimal input data vector. The attacker problem is defined as follows:
\begin{equation}\label{eq:maxmin}
Maximize\ \  \displaystyle \min_{d_i\in \mathcal{D}} \displaystyle \sum \limits_{a_j\in \mathcal{A}} U_{ij}\alpha_j
\end{equation}\vspace{-2mm}
\centerline{Subject to:}
\begin{equation}
\displaystyle \sum \limits_{a_j\in \mathcal{A}} \alpha_j=1
\end{equation}\vspace{-2mm}
\begin{equation}
\alpha_j \geq 0\ \forall a_j \in \mathcal{A}
\end{equation}
A maxmin strategy is one that maximizes the player's worst-case payoff. Here, the attacker tries to maximize the minimum impact of the Sybil attack by computing a probability distribution vector $\alpha$, according to which the attack will be distributed over the set of traffic controllers. The optimization problem is not linear but is equivalent to the following linear program:
\begin{equation}\label{prob1}
Maximize\ \  \rho
\end{equation}\vspace{-2mm}
\centerline{Subject to:}
\begin{equation}\label{equationrho}
\displaystyle \rho \leq \sum \limits_{a_j\in \mathcal{A}} U_{ij}\alpha_j\ \ \forall a_j\in\mathcal{A}
\end{equation}\vspace{-2mm}
\begin{equation}
\displaystyle \sum \limits_{a_j\in \mathcal{A}} \alpha_j=1
\end{equation}\vspace{-2mm}
\begin{equation}
\alpha_j \geq 0\ \forall a_j \in \mathcal{A}
\end{equation}
where $\rho$ is a variable defined for problem linearity. We try to maximize the value of $\rho$ while adhering to Constraint \ref{equationrho}. The problem can now be solved using the simplex method in polynomial time.\\
\indent When considering the problem from the perspective of the traffic control algorithm, the optimal defense strategy, represented by the probability distribution vector $\beta$, is derived by solving the following optimization problem. This is the minimax strategy that represents the mitigation approach.
\begin{equation}\label{defenseeq}
Minimize\ \  \displaystyle \max_{a_j\in \mathcal{A}} \displaystyle \sum \limits_{d_i\in \mathcal{D}}  U_{ij}\beta_i
\end{equation}\vspace{-2mm}
\centerline{Subject to:}
\begin{equation}
\displaystyle \sum \limits_{d_i\in \mathcal{D}} \beta_i=1
\end{equation}\vspace{-2mm}
\begin{equation}
\beta_i \geq 0\ \forall d_i \in \mathcal{D}
\end{equation}
This conservative strategy introduces potential resilience against Sybil attacks. This is equivalent to minimizing the largest attack impact that may occur in the worst case scenario. That is, the AMATSC's objective is to minimize the maximum attack impact (i.e., minimize the attacker's payoff). \\
\indent Based on the nature of traffic parameters usually exploited by current and emerging AMATSC algorithms, the vulnerability of these algorithms against data corruption attacks, particularly Sybil attacks, cannot be completely eliminated. Hence, we argue that an effective approach is to prepare the algorithm to deal with such attacks if ever materialized. Following the same mathematical transformations explained earlier, the nonlinear optimization problem becomes: 
\begin{equation}\label{prob2}
Minimize\ \  \phi
\end{equation}\vspace{-2mm}
\centerline{Subject to:}
\begin{equation}\label{equationphi}
\displaystyle \sum \limits_{d_i \in \mathcal{D}} U_{ij}\beta_i  \leq \phi \ \ \forall d_i \in\mathcal{D}
\end{equation}\vspace{-2mm}
\begin{equation}
\displaystyle \sum \limits_{d_i\in \mathcal{D}} \beta_i=1
\end{equation}\vspace{-2mm}
\begin{equation}
\beta_i \geq 0\ \forall d_i \in \mathcal{D}
\end{equation}
where $\phi$ is defined to make the problem linear, and should satisfy Constraint \ref{equationphi}. The two problems defined in Equations \ref{prob1} and \ref{prob2} are dual. The duality theorem explained in \cite{luenberger1973introduction} states that the maximum payoff that the TMC can achieve is equal to the minimum payoff that the attacker will receive. This is usually known as the ''value of the game'' and is achieved at the equilibrium point.

\section{Implementation Details}
\label{sec:4}
The lack of large-scale deployment of CAVs and technology limitations make theoretical analysis and simulation the main choices in the validation of our study. The realism of the simulation is thus a paramount aspect. To reproduce the road traffic network, we adopted the open-source microscopic traffic simulation package Simulation of Urban Mobility (SUMO) \cite{krajzewicz2010traffic}. 
SUMO has a Python Traffic Control Interface (TraCI) that interfaces it with an external application via a TCP socket connection. It permits SUMO to connect to other systems, such as the monitoring and control system. In fact, the TraCI interface allows the integration of Flow \cite{duan2016benchmarking}, a computational framework for deep RL and traffic control experiments.

\subsection{AMATSC Simulation Setup}

It has been proven that an RL  agent is capable of learning policies exceeding the performance of state-of-the-art traffic signal control programs \cite{mannion2016experimental}. 
Flow enables the use of a single agent to control an intersection with RL or a multi-agent algorithm to control a network of intersections (e.g. one controller agent acting on three intersections at the same time based on data collected from all three intersections). In this work, the implementation of MARL is contemplated as it is in this case that the benefits of learning should be most apparent and useful in the context of traffic signal control. RL-based AMATSC algorithms use different parameters as input to the RL algorithm. Among the data gathering techniques that assess the different parameters, we leverage the capabilities provided by the CAV technology within the traffic network and model them in Flow. Features are then extracted from the SUMO environment and provided as input to the control algorithm implemented in Flow. 

\indent Flow integrates SUMO with a standard deep RL library rllab, thereby permitting the training of large-scale RL experiments for traffic control tasks.
Also, Flow can be used to simulate CAV-based traffic control to enforce state changes in the traffic light program at the intersection. The changes in the traffic signal at intersections are dictated by the RL control component implemented in Flow, thus providing sophisticated, adaptive RL-driven traffic light programs. We used the configuration of the Proximal Policy Optimization (PPO) control algorithm, which is a policy-based RL method with significantly less computational complexity than other policy gradient methods \cite{schulman2017proximal}. 
We train the PPO RL agent under normal traffic conditions and obtain the resulting policy. 

\subsection{Evaluation Metrics}
\indent The mean trip waiting time is used as a performance metric to measure the impact of traffic control performed at each intersection compared to a baseline where static traffic programs are implemented. The waiting time represents the time in seconds during which the vehicle speed is below 0.1m/s. When RL agents are deployed, the waiting time of vehicles going through the network of controlled junctions decreases. In a stochastic traffic environment, in order to establish useful ranges for the waiting time metric, many simulation runs are performed and the mean trip waiting times are averaged. We implement several scenarios to assess the trip waiting time under normal traffic conditions, coordinated Sybil attacks, and when a mitigation solution is deployed.


\subsection{Baseline Scenario}
In the baseline scenario, traffic is flowing under normal traffic conditions. Moreover, the control logic implemented at each intersection is one of the types available in SUMO, which is either static or actuated (gap-based or delay-based). Vehicles will send a message to indicate their position and arrival time when they approach a traffic light. We identify three adjacent intersections in the transportation network, and collect the mean trip waiting time of the vehicles going through the identified intersections under this control logic. \\
\indent This scenario serves as the baseline for the evaluation of this metric. A MARL-based AMATSC managing the three intersections will aim at improving this metric. On the other hand, a coordinated Sybil attack on the network of intersections will try to degrade it. A repository reproducing the scenarios and results can be found at: \url{https://github.com/LiTrans/UrbanSybil/}. 

\subsection{MARL-based Control Scenario}
In this scenario, we run the experiment under normal traffic conditions with a trainable RL agent using the PPO algorithm to control the three intersections. 
The PPO algorithm requires the definition of \textit{state}, \textit{action} and \textit{reward} to train the traffic lights. The \textit{state} corresponds to the agent’s observation of the current environment, which includes the number of vehicles in each intersection observed as the closest to the traffic light. Particularly, the state space that is partially observed consists of the velocities, distances to the intersections, edge number from each direction, edge information, and traffic light states. The model uses the observation to compute the average velocity and flow density on each edge. \\
\indent The model also keeps a multi-dimensional array to keep track of how much time has passed since the last change to yellow for each traffic light, as well as to keep track of the flow direction and whether or not each traffic light is currently yellow. Results from SUMO are used by Flow for the generation of observations. The \textit{action} space specifies whether a traffic light is supposed to switch or not. The actions are sent to the traffic lights in the controlled network of intersections. The \textit{reward} represents the negative value of per vehicle delay minus a penalty for switching traffic lights. It is the reward associated with the previous state/action pair.\\
\indent The multi-agent environment provided by Flow enables the specification of the number of lights that the agent can observe (we set this number to three traffic lights in our simulation). The \textit{state} space in this context corresponds to the velocities, distances to intersections, edge number from each direction, and traffic light states of all three intersections. Hence, it represents the multi-agent shared model version of the network. The RL-based adaptive traffic signal controller issues an action for each traffic light agent. An agent receives a reward normalized by the number of lights. \\
\indent An experiment run in this scenario returns the trained RL model as a result. The final policy mapping is returned by the simulation. We then save the trained model to replay it by simply loading it. Since the algorithm requires a vectorized environment to run, we provide a reward of value 0 and an observation vector. The model is then able to predict the actions to take at each intersection of the network based on the observation provided. Thus, the model maps the states to actions to be performed. Actions from the policy are provided to the SUMO simulator. There is a minimum switch time implemented in the algorithm for each traffic light so that earlier RL commands are ignored. 

\subsection{Coordinated Sybil Attack Scenario}
The transportation network allows vehicles to frequently join and leave. This type of networks is susceptible to Sybil attacks, in which an attacker gains influence by joining a network under multiple colluding aliases. 
The traffic control algorithm exploits the data sent by the CAV in the range of an RSU to generate optimal signal plans. At all times, the RSU is fully aware of the number of vehicles by counting the number of messages received from both the Sybil and legitimate vehicles. The controller will consider any of the following metrics in the scheduling  algorithm: the number of vehicles, arrival time of each vehicle in each lane, vehicle speed, and vehicle delay, thus deriving other macroscopic metrics at the intersection. Consequently, every single vehicle within the RSU's communication range can potentially affect signal timing decisions.\\
\indent We simulate the coordinated Sybil attack in SUMO by injecting virtual vehicles. We vary the number of injected vehicles, the time and the duration of the injection. To measure the impact of the attack, trip waiting times of Sybil vehicles are not considered in the computation of the mean trip waiting time metric because they simply do not exist. Thus, we consider only the trip waiting times of real vehicles and compare them to the baseline scenario, where traffic is flowing under normal traffic conditions e.g. no attack on AMATSC. 
In our coordinated Sybil attack implementation, the attacker deploys Sybil vehicles per each RSU. Thus, the network of three intersections controlled by the RL algorithm is targeted with different Sybil arrival rates at every intersection. 
 To reflect a generic attack scenario where the attacker is not necessarily intelligent, we implement an attack strategy where the attacker targets the critical phase that causes the largest delay, instead of implementing the best-case attack scenario generated by our game. Hence, each traffic signal in the network is identified as a potential attack target. The measured attack impact was substantial even when the optimal attack policy was not adopted. Nevertheless, our mitigation approach is designed to deal with the worst-case attack scenario.

\begin{figure}[!t]
\begin{center}
  \includegraphics[scale=0.7]{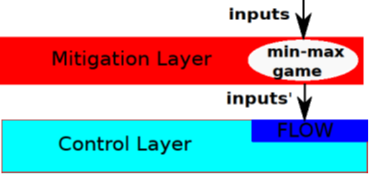}
  \captionsetup{width=.88\linewidth}
  \caption{The mitigation layer is implemented on top of the control layer for increased application resilience.}
  \label{fig:layers}
\end{center}\vspace{-2mm}
\end{figure}

\subsection{Threat Mitigation Scenario}
Here, we implement the minimax game as a mitigation layer on top of the control layer. Figure \ref{fig:layers} illustrates the defense strategy as a layer of protection against coordinated data corruption attacks on the AMATSC application regardless of intrusion detection performance. To reduce the attack impact, this layer will adjust the inputs provided by SUMO to the AMATSC algorithm implemented in Flow. First, we need to compute the maximum flow on the 12 traffic segments of the intersections controlled by the RL agent (three junctions, each having 4 segments).

\begin{figure}[!t]
\centering
	\includegraphics[scale=0.34]{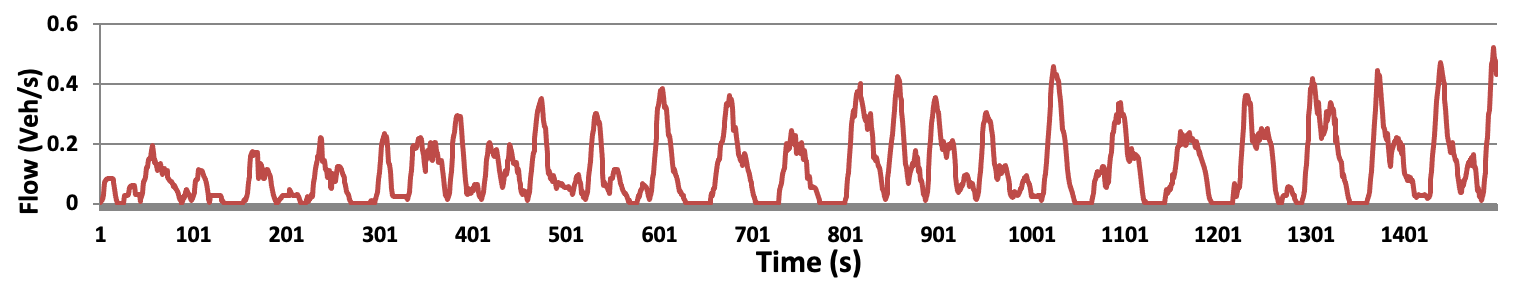}
	  \captionsetup{width=.88\linewidth}

\caption{Variation of traffic flow on an urban road network.}
\label{fig:Flowedge}
\end{figure}

\begin{figure}[t]
	\includegraphics[scale=0.4]{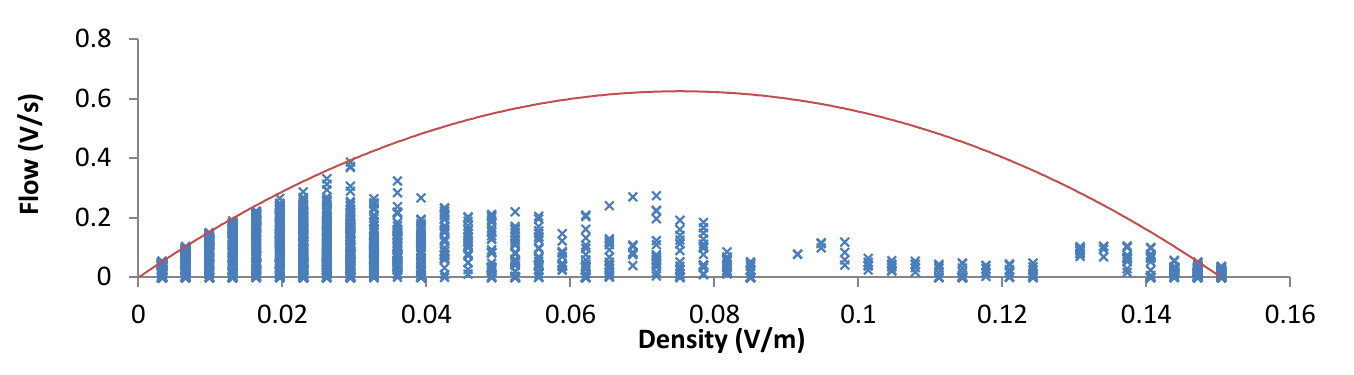}
	\centering   \captionsetup{width=.88\linewidth}

	\caption{Flow-Density relationship on an edge in the urban baseline scenario.}
       \label{fig:flowdensity}
\end{figure}

 Figure \ref{fig:Flowedge} shows the variation of traffic flow on an urban road network. Macroscopic stream models represent how the behavior of one parameter of traffic changes with respect to another. Precisely, the relation between the flow and density of an edge is shown in Fig. \ref{fig:flowdensity}. Also, the assumed linear equation between speed and density is given by:
\begin{equation}
v = v_{f} - \Big(\frac{v_{f}}{k_{j}}\Big)\times k
\end{equation}
where $v$ is the mean speed at density $k$, $v_{f}$ is the free speed, and $k_{j}$ is the jam density. The relation between flow and density can be derived using $q=k\times v$, thus getting the following parabolic equation:\vspace{-2mm}
\begin{equation}\vspace{-1mm}
q = v_{f}\times k - \Big(\frac{v_{f}}{k_{j}}\Big)\times k^2
\end{equation} 
Finally, we derive the relation between speed and flow: 
\begin{equation}\vspace{-1mm}
\label{eq:speedflow}
q = k_{j} \times \Big(\frac{v - v^2}{v_{f}}\Big) 
\end{equation}

Once the relationship between the fundamental variables (density, flow, speed) of traffic flow is established, the boundary conditions can be derived. The one of interest is the maximum flow. From Equation (\ref{eq:speedflow}), we find the critical density at the maximum flow using the following derivative:\vspace{-1mm}
\begin{equation}\vspace{-2mm}
\frac{dq}{dk} = v_{f}\times \Big[1 - 2\times \Big(\frac{k_{c}}{k_{j}}\Big)\Big]=0
\end{equation} 
\begin{equation}\vspace{-1mm}
k_{c} = \Big(\frac{k_{j}}{2}\Big)
\end{equation}

Therefore, the density corresponding to maximum flow can be approximated by half the jam density. Once we get \begin{math}k_{c}\end{math}, we can derive the maximum flow, denoted by \begin{math}q_{max}\end{math}. \vspace{-2mm}
\begin{equation}\vspace{-1mm}
\label{eq:qmax}
q_{max} = \Big(\frac{v_{f} \times k_{j}}{4}\Big) 
\end{equation}
Thus, the maximum flow is approximated by one fourth the product of free flow and jam density. 
In the second part of the scenario, we extract from the simulation the current flow observed on the segment of interest and compare it to the maximum flow:\vspace{-2mm}
\begin{equation}\vspace{-1mm}
\Delta = q_{max}(t) - q_{actual}(t)
\end{equation} 

From the perspective of the attacker, the aim is to inject Sybil vehicles on the edges that best suit the attack's goal according to the current observations of density and flow. Particularly, the attacker will consider the edges where density is low to medium and the maximum injection rate will be dictated by $\Delta$. Knowing that the RL agent keeps track of the average velocity of the vehicles observed, the attack will aim to decrease the average speed, thus inducing the model to propose suboptimal actions. The minimax game mitigates the attack by limiting the influence the attack achieves through the Sybil nodes. 

To this end, this scenario consists of the implementation of the defense optimisation problem in Equation \ref{defenseeq}. The output of the proposed minimax game is the vector $\beta$ corresponding to the weights of information collected on each edge (e.g., its reliability). Such high-level weights could be interpreted in a variety of ways depending on the traffic control module under investigation. This is one of the advantages of our proposed game model, which offers a relative level of flexibility in how the mixed strategy of mitigation actions would be actually implemented. In our simulation, each element of the vector corresponds to the percentage of vehicles to be considered on each edge at each iteration $t$ by the RL agent for the computation of the input average speed. These values represent the modified inputs fed to the control layer as the new state observed by the MARL agents, which is the most fundamental component of our resilience-by-design scheme.

\begin{figure*}[!t]
    \centering
    \begin{subfigure}[b]{.45\linewidth}
        \centering

        \includegraphics[width=0.8\textwidth]{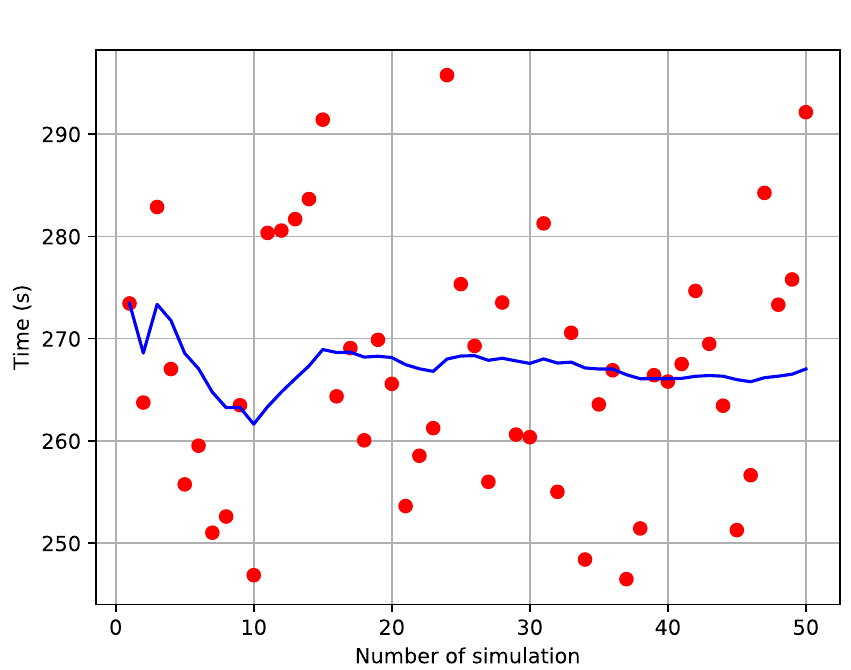}
\captionsetup{width=.9\linewidth}	
        \caption{\footnotesize Baseline scenario: Mean trip waiting time in normal conditions on Junction 386.}\label{MTLresults_1}
    \end{subfigure}\vspace{2mm}
\quad
    \begin{subfigure}[b]{.45\linewidth}
        \centering

        \includegraphics[width=0.8\textwidth]{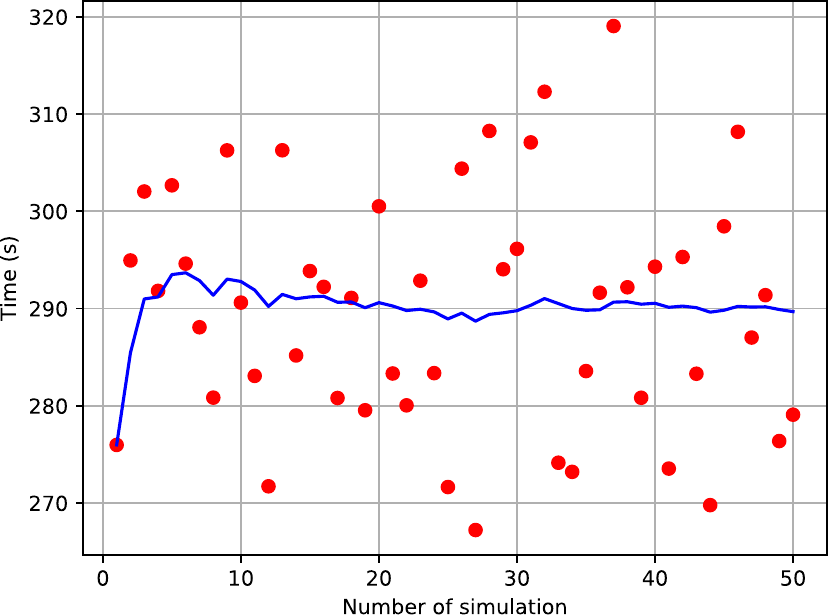}	
\captionsetup{width=.9\linewidth}
        \caption{\footnotesize Sybil attack scenario: Mean trip waiting time under attack on Junction 386.}\label{MTLresults_2}
    \end{subfigure}\vspace{2mm}

    \begin{subfigure}[b]{.45\linewidth}
        \centering

        \includegraphics[width=0.8\textwidth]{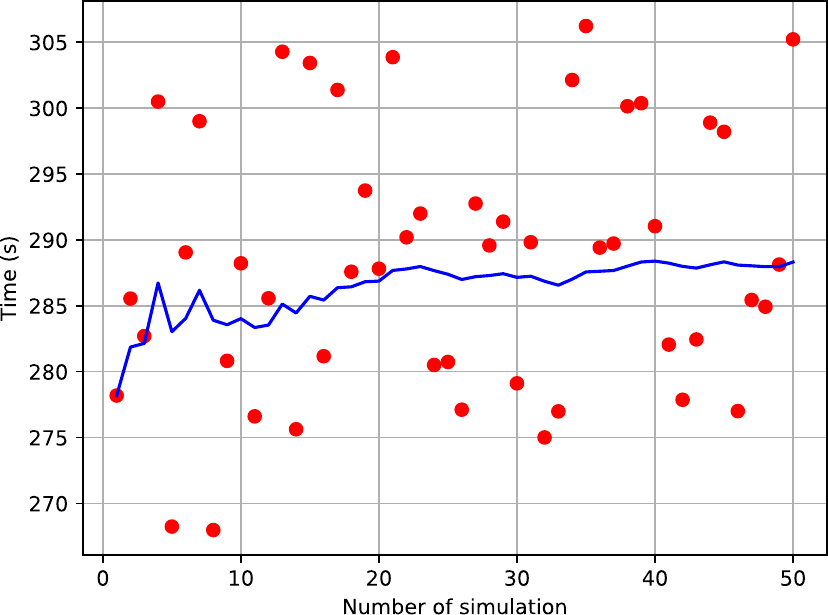}	
\captionsetup{width=.9\linewidth}
        \caption{\footnotesize Baseline scenario on networked junctions 386, 428, and 376: Mean trip waiting in normal conditions.}\label{MTLresults_3}
    \end{subfigure}
\quad
    \begin{subfigure}[b]{.45\linewidth}
        \centering

        \includegraphics[width=0.8\textwidth]{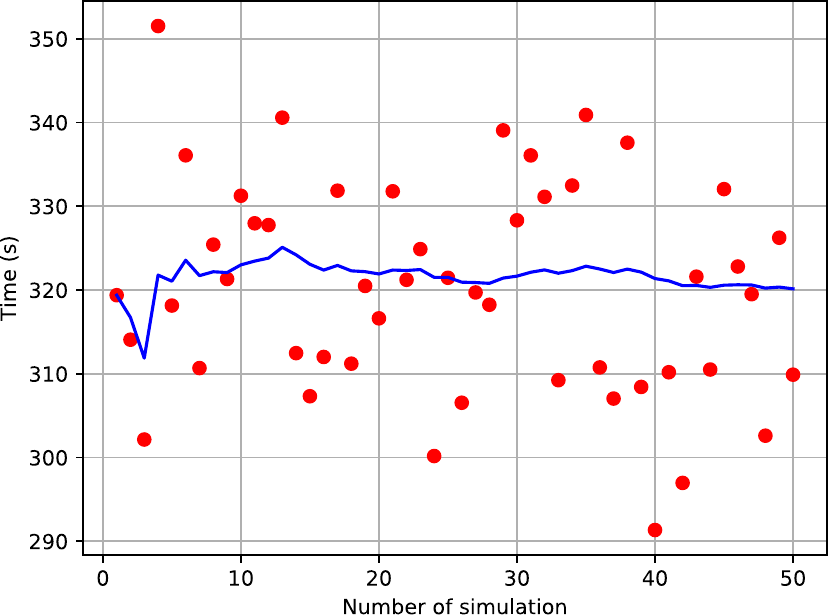}	
\captionsetup{width=.9\linewidth}
        \caption{\footnotesize Sybil attack scenario on networked junctions 386, 428, and 376: Mean trip waiting time under attack.}\label{MTLresults_4}
    \end{subfigure}
    \captionsetup{width=.9\linewidth}

	\caption{Results of the baseline and coordinated Sybil attack scenarios on a single junction and three junctions.}
\end{figure*}

\section{Experimental Results}
\label{sec:5}
To demonstrate that the discovered security flaws have high practical implications, this section assesses the impact of the coordinated Sybil attack  on a realistic transport network and provides the results of the mitigation strategy, which can be deployed to protect existing and emerging ATSC systems.

\subsection{Impact of Coordinated Sybil Attacks}

We perform the attack on the real urban network of the city of Montr\'eal using real-world traffic data and signal timings.

\subsubsection{Dataset Description}
Information on traffic demand in Montr\'eal was obtained in the form of an Origin-Destination matrix (O-D matrix) from the Survey of the Montr\'eal Metropolitan Area carried out in 2013 by the provincial government. The zone corresponds to the city downtown and occupies an approximate area of 4 $km^2$. We also used the information on the traffic signal plans of the modeled network for the period of 8AM to 9AM, provided by the Transport Division of the City of Montr\'eal. The dataset does not include trips of pedestrians, cyclists, public transport (buses), nor commercial vehicles (trucks). The traffic conditions of the Montr\'eal road network were experimentally reproduced under normal conditions in SUMO as per the baseline scenario. The real road network of Montr\'eal was also used in the coordinated Sybil attack scenario.  

\subsubsection{Results and Analysis}
The realistic evaluation of the impact of the coordinated Sybil attack in terms of mean trip time loss demonstrates the potential physical impact of real cyber-physical attacks, which exploit the perception layer to target the application layer. We validate that there is no need to rely on assumptions of the effect of attacks against the traffic control components and that our mitigation has solid merit in improving the resilience of these components. We provide results for both the baseline and Sybil attack scenarios on one junction and three networked junctions of downtown Montr\'eal. 

We identified the 50 most critical traffic lights based on the average mean time loss in several simulation runs. Fig. ~\ref{MTLresults_1} shows the mean trip waiting time of the most critical junction. This junction is controlled by a gap-based actuated traffic control. This control scheme works by prolonging traffic phases whenever a continuous stream of traffic is detected. It switches to the next phase after detecting a sufficient time gap between successive vehicles. This allows for better distribution of green-time among phases and also affects cycle duration in response to dynamic traffic conditions. We considered the maximum time gap to be of three seconds between successive vehicles that will cause the current phase to be prolonged. The minimum duration and max duration of green-time are 5 and 45 seconds, respectively.

Fig. \ref{MTLresults_2} shows the impact of Sybil attack on the same junction as the baseline scenario. Sybil vehicles were injected around the junction long enough to be detected by the RSU, which then receives the wrong information about the state of traffic flow. The Sybil attack increased the mean trip time loss by approximately 23 seconds compared to the baseline.

\indent Since AMATSC is not yet deployed in the transportation network of the City of Montr\'eal, Fig. \ref{MTLresults_3} shows the average mean trip waiting time of vehicles on three neighboring but isolated intersections under the baseline scenario. In the next section, we demonstrate that, if the junctions were interconnected and AMATSC was implemented, the performance in terms of trip time loss is improved. Fig. \ref{MTLresults_4} shows the impact of the Sybil attack on the same three junctions. The junctions are also controlled by a gap-based actuated traffic control. We notice that the trip time loss under attack increased by 34 seconds, that is, approximately by up to 50\% more than the Sybil attack targeting a single junction. 

In \cite{yen2018falsified}, they presented a study on the impact of time spoofing attacks on different traffic signal control algorithms in single and multiple intersections under both homogeneous and heterogeneous arrivals. They showed that the delay-based scheme is more vulnerable to time spoofing attacks compared to the sum-of-delay-based scheme. In addition, the hybrid scheme that combines delay-based and queue-based performs similarly to the queue-based scheme when under attack. From their conclusions about the time spoofing attack, we can infer the same about our coordinated Sybil attack. We delivered evidence of the damage that can be caused in terms of increasing the time loss under a gap-based control algorithm. It will be further demonstrated that even more substantial damage can be caused under other scheduling algorithms, such as AMATSC.


\subsection{Attack Mitigation Performance}
\begin{figure}[!t]
\centering
	\includegraphics[scale=0.35]{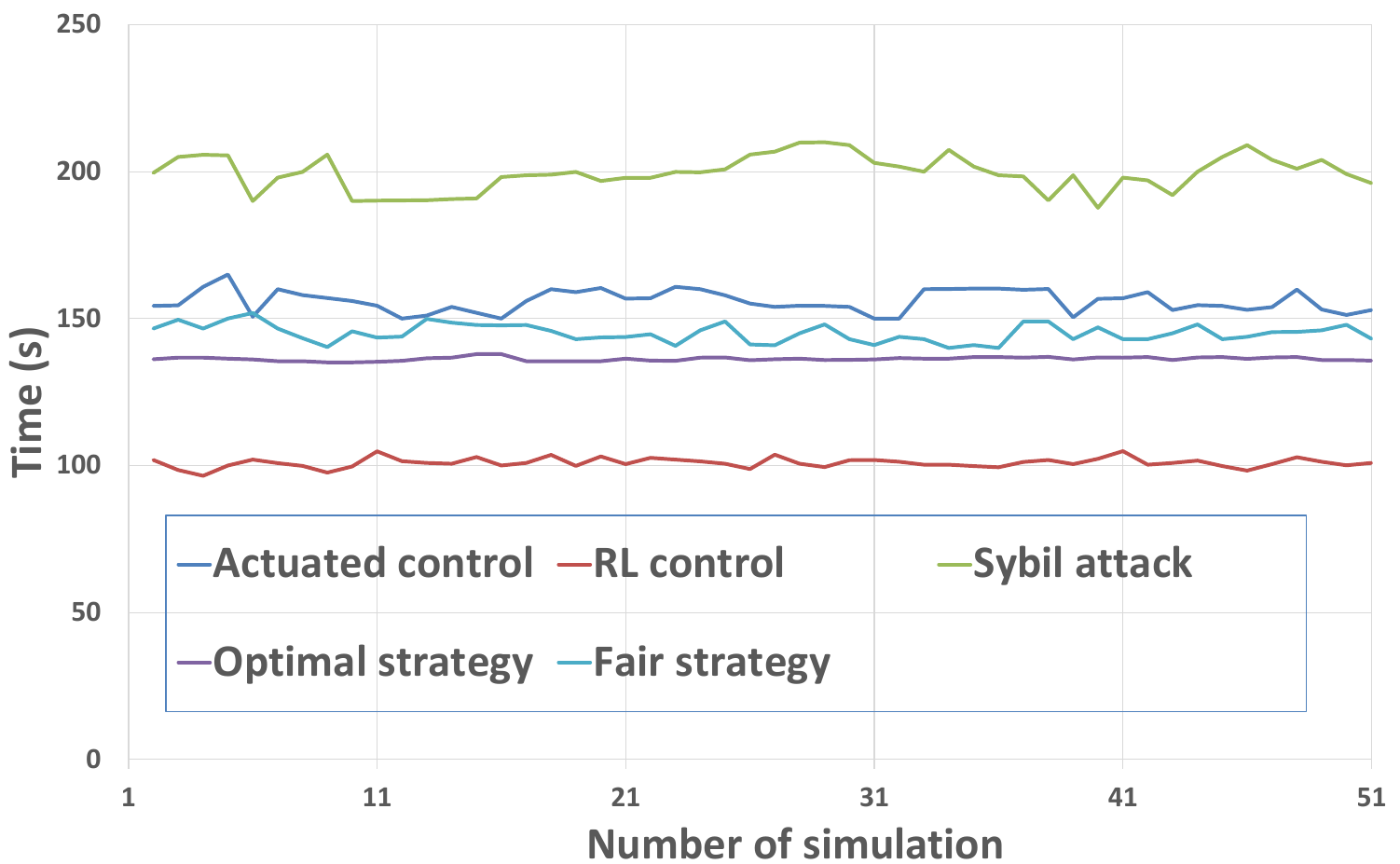}
	\centering
	\caption{Grid network scenarios: Mean trip waiting time.}
       \label{Mitigationresults}
\end{figure}

The mitigation scenario described earlier is simulated on a  grid network topology. We use the pre-built configurable traffic light grid environment implemented in Flow to conduct the simulations. We customized the environment/network parameters as per TABLE \ref{topoFLOW}. Using the 10x10 grid network topology, we simulate the baseline scenario under normal traffic conditions, where the traffic light logic exerts control over individual traffic lights using SUMO actuation. In fact, the default SUMO actuated traffic lights are fine-tuned on many iterations with varying parameters of phase duration and state. Fig. \ref{Mitigationresults} shows the mean trip waiting time around three junctions of the grid network for different scenarios. The dark blue curve corresponds to actuated control and represents the baseline scenario used for the comparison with the other scenarios. The mean trip waiting time is 154.3 seconds. 

The red curve in Fig. \ref{Mitigationresults} shows the results for the same environment but with three traffic lights controlled by an RL agent. The other traffic lights remain controlled by the default SUMO actuated type of traffic control. This is where we switch from the non-RL to the MARL experiment in Flow to apply RL-specified traffic light actions via TraCI. Precisely, to simulate AMATSC on a network of intersections, we implement the multi-agent experiment where agents use the same policy.\\
\indent We set the state to be partially observable. In fact, on each edge around the controlled intersection, we want to know the velocity of every vehicle. For each traffic light, we want to know its current state, i.e., what direction it is flowing, when it last changed, and whether it was yellow. This subset of information is then provided to the controller. Each agent is considered as a single intersection controlling its traffic lights. The observation space is defined as normalized velocities of nearby vehicles, for every intersection. We set the action space to be discrete (e.g. the action space specifies whether a traffic light is supposed to switch or not) and to directly match the number of traffic intersections to be controlled. Because of the shared policy, instead of computing the actions, state, and reward for a single agent, as a reward, the RL-based controller will penalize the large delay and boolean actions that indicate a switch (with the negative sign) for all the controlled agents in the network. The same policy is used by each agent, and actions from the policy are provided to the SUMO simulator. \\
\indent We notice from Fig. \ref{Mitigationresults} that MARL-driven control is able to decrease the mean trip waiting time by approximately 52.4 seconds compared to the baseline scenario on three actuated traffic lights under normal conditions. We then run the coordinated Sybil attack scenario on the grid network with three traffic lights controlled by an RL agent and we present the results in Fig. \ref{Mitigationresults}. The green curve shows the attack's impact on adaptive traffic control. The increase of approximately 97.7s compared to the RL-based control scenario motivates the need for mitigation, which proves to be required to limit the impact of this sophisticated attack on AMATSC.\\
\indent In Fig. \ref{Mitigationresults}, the purple curve presents the results of the optimal mitigation strategy when the network of intersections is the target of a coordinated Sybil attack. We are able to improve time loss by approximately 48.9\%. Our design proves to mitigate the attack by limiting the influence the attacker gains through Sybil nodes.  Moreover, we realize that the optimal strategy performs better than the baseline, which validates the fact that we can't just ignore the link that has an attack for a certain time or minimize its effect by switching to fixed time plan control or actuated control instead of adaptive controller without applying our mitigation strategy. On the other hand, the performance of RL-based control when the mitigation approach is applied (purple curve) would still be degraded compared to its performance under normal settings (red curve). As expected, the mitigation solution is designed to attenuate the attack's impact, not to eliminate it completely. \\
\indent Fig. \ref{Mitigationresults} compares the performance of the optimal mitigation strategy with respect to the fair strategy wherein the mitigation does not consider the probability distribution vector $\beta$ generated by the minimax game solution, and which corresponds to the percentage of vehicles to be considered in the algorithm's input on each edge as part of the mitigation process. Instead, the fair strategy modifies the weighted speed values on each edge by taking the same number of vehicles present on each edge regardless of which edge might be more impacted by the coordinated Sybil attack. This strategy does not account for the potential presence of coordinated Sybil vehicles within the network, which attempt to achieve a high-impact attack by optimizing their presence on the networked traffic edges. Hence, this strategy is not supposed to provide optimal mitigation, simply because it does not ponder the payoff matrix that we created in Equation \ref{payoff} and is not prepared to deal with the worst-case attack scenario. \\
\indent In our simulation, the fair strategy considered half of the number of vehicles present on each edge of the intersections controlled by the RL agent for the computation of the average speed on that edge at each iteration. The light blue curve shows that the fair strategy improves time loss incurred under a coordinated Sybil attack by approximately 26.5\%. As expected, the fair strategy does not perform as good as the optimal strategy, and thus is not appropriate to implement as part of a resilience-by-design defense approach.

\begin{table}[!t]
	\centering
\caption{Grid network topology.}
	\resizebox{0.35\textwidth}{!}{%
	\renewcommand{\arraystretch}{1.5}
	\begin{tabular}{|c|c|}
    \hline
    Horizon & 5000 seconds\\
    \hline
    Number of rows & 10 \\
    \hline
    Number of columns & 10 \\
    \hline
    Dimensions & 5500mx4800m \\
    \hline
    Lanes & Horizontal:2, Vertical:2 \\
    \hline
    Max gap & 3.0  \\
    \hline
    Detector gap & 0.8  \\
    \hline
    Speed limit & 35 m/s\\
    \hline
    Inflows & top:20 bottom:40 left:40 right:50 \\
\hline
\end{tabular}}\vspace{-2mm}
	\label{topoFLOW}
\end{table}

\begin{figure}[!t]
\centering
	\includegraphics[scale=0.35]{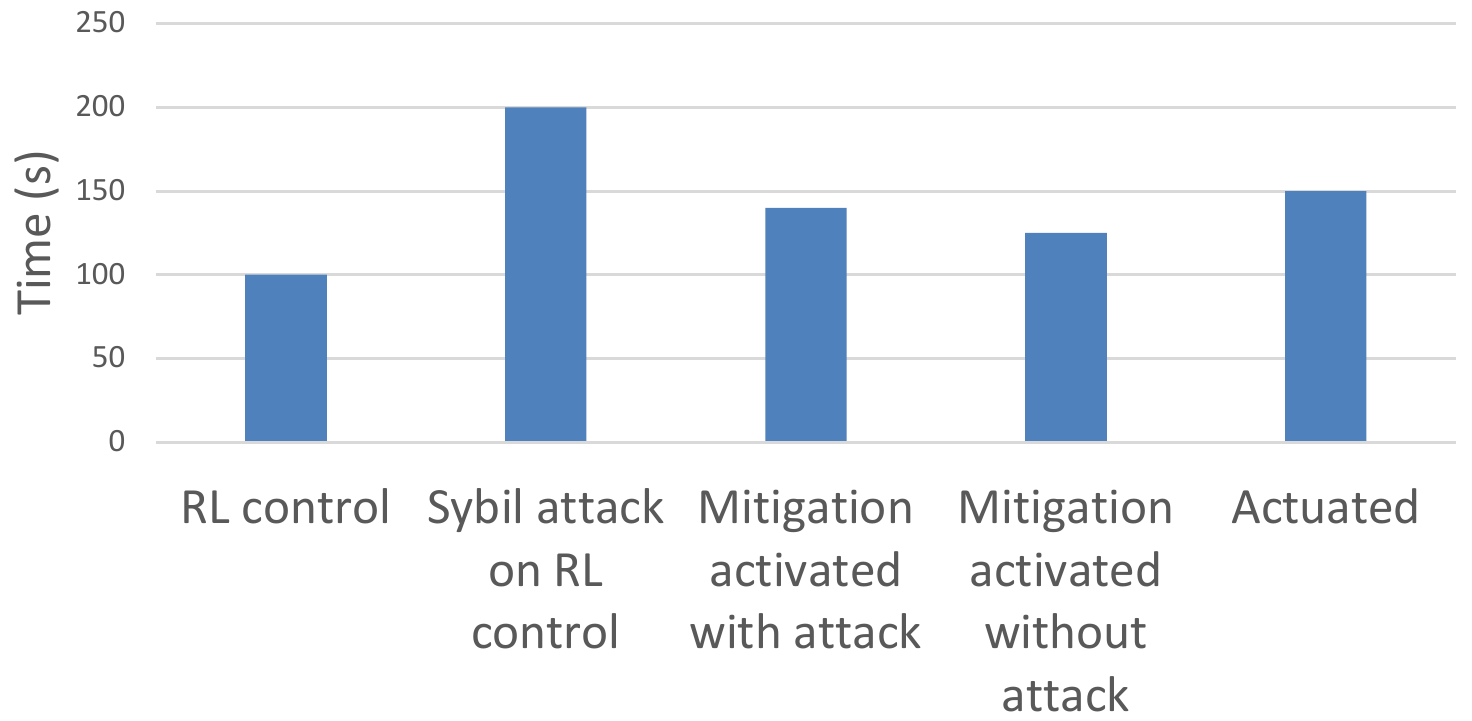}
	\centering
	\caption{Performance in terms of mean trip wait time of the mitigation under different scenarios.}
       \label{Mitigationresults2}
\end{figure}

\indent Fig. \ref{Mitigationresults2} presents the results when the mitigation approach is activated without the Sybil attack compared to the other scenarios. Compared to the RL control under no attack, the performance was slightly degraded of approximately 25\% but is still better than the actuated control performance, which justifies the activation of our mitigation approach even in absence of attack.  

\section{Discussion}
We have proven that the threat posed by sophisticated attacks on traffic control systems is of particular concern. To counter this threats, the first line of defense is to implement detection solutions. However, most of state-of-the-art Sybil attack detection approaches have serious shortcomings. To be effective, a detection approach requires precise knowledge of the actual system being controlled and the control discipline employed. Traffic control systems are highly vulnerable to cyber-physical attacks, and domain-agnostic security solutions are too generic to be able to comprehend and detect intelligent, sophisticated attacks. In critical infrastructures such as ITS, the attack can cause physical, disruptive damage. Therefore, a proactive defense solution is required, without necessarily relying on attack detection. This paper proposes an effective mitigation strategy that can be integrated into the application layer, providing resilient-by-design AMATSC systems. 

To maintain an optimal performance of the RL-driven traffic control, the proposed mitigation approach is not required to be deployed at all times. It can be adopted at the application layer when the control logic has suspected an attack and requires to be proactively prepared to face its impact on decision making. The proposed mitigation solution constitutes an invaluable line of defense to integrate within a defense-in-depth strategy in highly vulnerable environments running in adversarial settings. It is the first step towards the design of an adaptive cyber-physical security approach tailored to traffic control systems \cite{tsigkanos2016interplay}. Also, one of the highlights of our mitigation solution is that it does not rely on the detection and identification of Sybil nodes to mitigate the attack impact, unlike most of existing solutions.

\section{Conclusion and Future Work}
\label{sec:6}
A new and emerging challenge to the design of AMATSC algorithms is explored. We stress the fact that the implementation of the traffic control logic should be attack-aware. We present a new, highly realistic threat model, namely, coordinated Sybil attack, that targets a network of intersections controlled by adaptive control algorithms to sabotage their decision making. We implement the attack and investigate its impact using a real traffic dataset under real-world intersection settings. The obtained results as well as the exposed vulnerabilities of emerging AMATSC systems should be taken seriously when pondering future design choices and implementations.

To respond to anticipated intelligent data corruption attacks against AMATSC, we present a mitigation strategy as a layer of protection in case of detection failure. The devised minimax game model enables the AMATSC algorithm to take optimal decisions under attack.  The optimal mitigation strategy showed a substantial improvement of time loss by approximately 48.9\%. It can be potentially integrated at the application layer as part of a resilience-by-design approach.

In the future, we will study the impact of our coordinated Sybil attack on different traffic control algorithms with regards to the metrics used. Some AMATSC algorithms may be more vulnerable than others to our attack. Hybrid schemes that combine metrics may perform better when under attack than schemes that use single metrics. Similarly, the effectiveness of the mitigation may vary depending on the deployed control algorithm, which requires further investigation.  


  \bibliographystyle{ieeetr}
  \bibliography{ref}

\renewenvironment{IEEEbiography}[1]
{\IEEEbiographynophoto{}}
{\endIEEEbiographynophoto}
\vspace{-40pt}
\begin{wrapfigure}{l}{0.15\textwidth}
	\centering
	\includegraphics[width=0.15\textwidth,height=3cm]{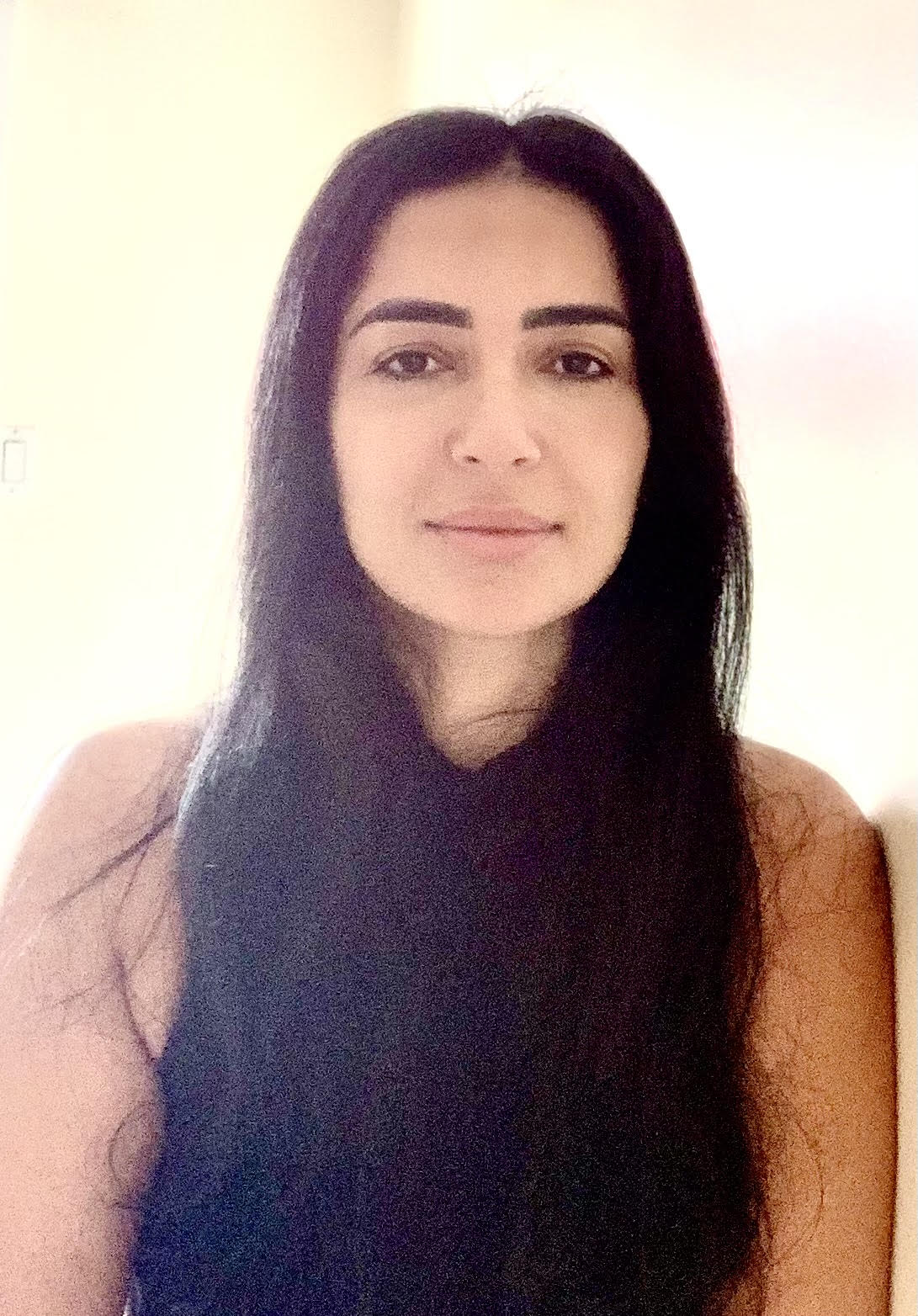}
	\centering
\end{wrapfigure}
\begin{IEEEbiography}
	\scriptsize
	\textbf{Ranwa Al Mallah} received her Ph.D. degree in Computer Science from Polytechnique Montreal. She is currently an assistant professor in cybersecurity at the Royal Military College of Canada. Previously, she was a postodoctoral fellow at Ryerson University.  She was also a member of the SecSI Research Laboratory at Polytechnique Montreal. 
	Her current research goal is to develop multidisciplinary, secure and highly intelligent solutions for the planning, design and operation of cyber physical systems in the context of smart cities.
\end{IEEEbiography}
\vspace{-10pt}

\begin{wrapfigure}{l}{0.15\textwidth}
	\centering
	\includegraphics[width=0.16\textwidth,height=3.5cm]{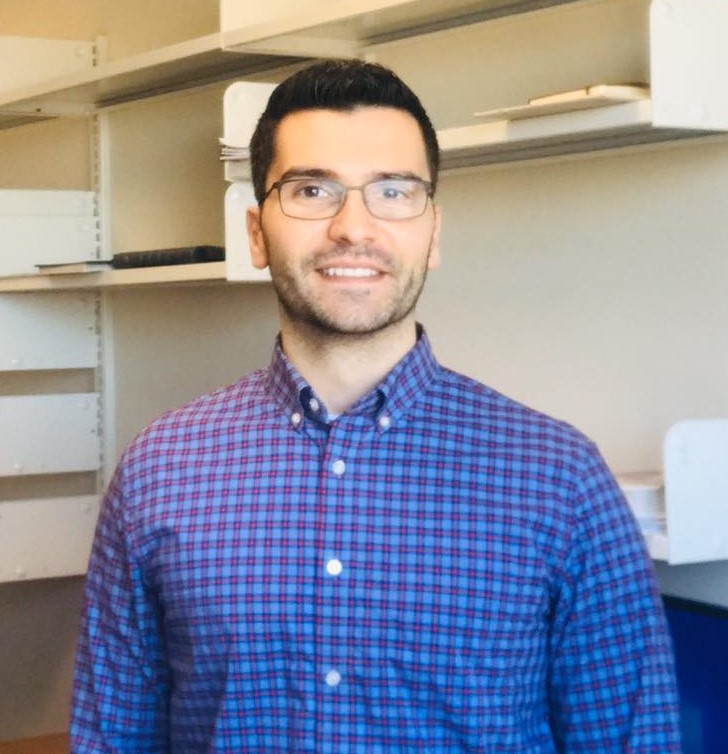}
	\centering
\end{wrapfigure}
\begin{IEEEbiography}
	\scriptsize
	\textbf{Talal Halabi} is an assistant professor in Applied Computer Science at the University of Winnipeg. Previous to that, he served as a postdoctoral fellow at Queen's University in Kingston, Canada. He received his Ph.D. in Computer Engineering in 2018 from Polytechnique Montréal at the University of Montreal. He has a M.Res. in Telecommunications from Saint Joseph University of Beirut, and a M.Eng. in Electrical and Electronics Engineering from the Lebanese University. His research interests are in the areas of security and resilience of Cyber-Physical Systems and the Internet of Things (IoT).
\end{IEEEbiography}

\begin{wrapfigure}{l}{0.15\textwidth}
	\centering
	\includegraphics[width=0.16\textwidth,height=3.5cm]{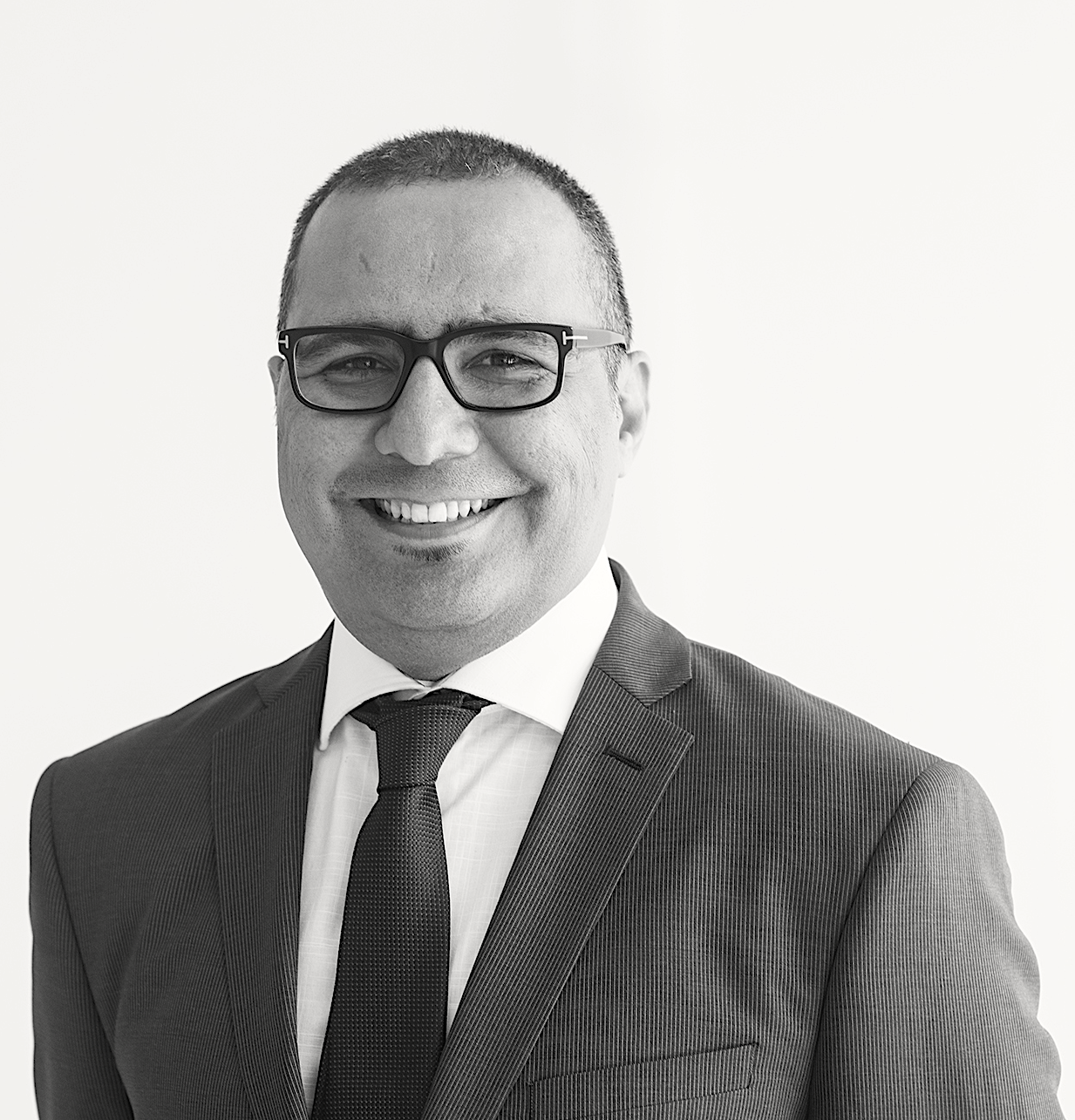}
	\centering
\end{wrapfigure}
\begin{IEEEbiography}
	\scriptsize
\textbf{Bilal Farooq} received B.Eng. degree (2001) from the University of Engineering and Technology, Pakistan, M.Sc. degree (2004) in Computer Science from Lahore University of Management Sciences, Pakistan, and Ph.D. degree (2011) from the University of Toronto, Canada. He is the Canada Research Chair in Disruptive Transportation Technologies and Services and an Associate Professor at Ryerson University, Canada. He received Early Researcher Award in Québec (2014) and Ontario (2018), Canada. His current work focuses on the network and behavioral implications of emerging transportation technologies and services.
\end{IEEEbiography}

\end{document}